\documentclass[preprints,article,accept,pdftex,oneauthor]{Definitions/mdpi} 
\usepackage{bm}
\renewcommand{\vec}[1]{\bm{#1}}

\usepackage{xspace}

\usepackage{soul}
\usepackage{etoolbox}

\robustify\hl
\colorlet{bckgr}{white!100!}
\sethlcolor{bckgr}
\newcommand{\Tm}{Tm~\textsc{ii}\xspace}
\firstpage{1} 
\pubvolume{1}
\issuenum{1}
\articlenumber{0}
\pubyear{2025}
\copyrightyear{2025}
\datereceived{ } 
\daterevised{ } % Comment out if no revised date
\dateaccepted{ } 
\datepublished{ } 
\hreflink{https://doi.org/} % If needed use \linebreak
\Title{Calculation of hyperfine structure in \Tm}

\TitleCitation{Calculation of hyperfine structure in \Tm}

\Author{Andrey I. Bondarev $^{1,2}$\orcidA{}}

\AuthorNames{Andrey I. Bondarev}

\isAPAStyle{%
       \AuthorCitation{Bondarev, A. I.}
         }{%
        \isChicagoStyle{%
        \AuthorCitation{Bondarev, Andrey I.}
        }{
        \AuthorCitation{Bondarev, A. I.}
        }
}

\address{%
$^{1}$ \quad Helmholz-Institut Jena, 07743 Jena, Germany; a.bondarev@hi-jena.gsi.de \\
$^{2}$ \quad GSI Helmholtzzentrum f\"ur Schwerionenforschung GmbH, 64291 Darmstadt, Germany}

\abstract{
The first measurements of the magnetic dipole hyperfine structure constants $A$ in singly ionized thulium revealed substantial discrepancies with the corresponding
calculations [Mansour et al., NIMB \textbf{40}, 252 (1989)]. More recent measurements [Kebapcı et al., ApJ \textbf{970}, 23 (2024)] expanded \hl{the} very limited dataset of that work and demonstrated that two of the previously reported experimental $A$ values were incorrect. \hl{This motivates to perform new theoretical calculations.} In this work, we employ the configuration interaction method to calculate the $A$ constants for several low-lying levels in \Tm, with the random-phase-approximation corrections also taken into account. Our results show good agreement with the new experimental data and provide reliable predictions for additional states where measurements are not yet available.
}
\begin{document}
%%%%%%%%%%%%%%%%%%%%%%%%%%%%%%%%%%%%%%%%%%
\section{Introduction}
%%%%%%%%%%%%%%%%%%%%%%%%%%%%%%%%%%%%%%%%%%
Thulium is located near the end of the lanthanide series in the periodic table of the elements \hl{and has only one stable isotope, $^{169}_{\;\;69}$Tm, with} nuclear spin $I=\tfrac{1}{2}$. Thus, the hyperfine interaction of electrons in \hl{its} ions with the nuclear moments is determined solely by the magnetic dipole contribution, while the electric quadrupole contribution is absent. The resulting hyperfine structure (HFS) of the electronic levels is therefore characterized exclusively by the magnetic dipole HFS constant~$A$.

Singly ionized thulium, \Tm, as a system with an open $f$ shell, has a complex and dense spectrum, which was studied experimentally~\cite{Martin78,Wyart11} and theoretically~\cite{Quinet99,Radžiūtė21}. Although the experimental research on transition rates and lifetimes is sufficiently extensive~\cite{Blagoev94,Anderson96,Wickliffe97,Rieger99,Xu03,Tian16,Wang22,DenHartog24} (see Refs.~\cite{Quinet99,Radžiūtė21} for theoretical predictions), only two papers report measurements of the HFS, and results of only \hl{one} corresponding calculation have been published.
%
%is relevant for astrophysics. (abundances)
%Considering the current state of research, reliable atomic data that will enable accurate determination of rare earth element abundances for the transitions under study are as important as obtaining high-resolution and high-SNR ratio stellar spectra.
%
\citet{Mansour89} used collinear fast-ion-beam laser spectroscopy to determine HFS constants $A$ for eight levels. Their measurements of the $A$ constants for the four lowest states were compared with the results of the  multiconfiguration Dirac-Hartree-Fock (MCDHF) calculations %performed by K.T.~Cheng 
(see Ref.~\cite{Cheng85} for calculation details). They found a clear discrepancy between the experimental data and theoretical predictions for the $(4f^{13}(^2F_{5/2})6s_{1/2})_2^o$ level and concluded that configurations containing an open $s$ shell are in substantial disagreement with the calculations.
Recently, \citet{Kebapcı24} reported $A$ constants extracted from the hollow-cathode-discharge-lamp emission spectra recorded by a Fourier transform spectrometer. Parts of the spectra in other wavelength ranges had previously been analyzed to study HFS and transition rates in neutral thulium~\cite{Parlatan22,Kebapcı22,Bondarev24}.  
They determined 27 additional $A$ constants beyond those reported by~\citet{Mansour89} and showed that two of the values in that work, including the one for the $(4f^{13}(^2F_{5/2})6s_{1/2})_2^o$ level, were incorrect. Their revised value for this level partially resolves the discrepancy with the MCDHF result reported in Ref.~\cite{Mansour89} and motivates new calculations.

Meanwhile, an experimental campaign at CERN aimed \hl{to measure} nuclear moments and changes in root-mean-square radii in radioactive thulium isotopes is ongoing~\cite{Cheal:2834596,Cheal:2872390,Cheal:2912229}. As a groundwork for these experiments, the hyperfine structure and transition rates in the stable isotope are also being studied at COALA, an offline collinear laser spectroscopy setup for high-precision measurements at TU Darmstadt (Germany)~\cite{Konig.2020b}. This technique enables the determination of the $A$ constants with significantly higher accuracy than the approaches mentioned above.
\hl{Furthermore, a spectroscopic investigation of \Tm} \hl{focused on establishing its use as a platform for advanced quantum applications is currently underway at the University of California, Los Angeles}~\cite{Müller25}.

In this study, we calculate the magnetic dipole HFS constants $A$ and Land\'e $g$ factors for the low-lying levels of the $4f^{13}6s$ and $4f^{13}5d$ configurations in \Tm using a configuration interaction (CI) method with the random-phase approximation (RPA) included. Our results show good agreement with the available measurements and provide predictions for states where experimental data are not yet available. \hl{Atomic units} $\hbar=e=m_\mathrm{e}=4\pi\varepsilon_0=1$ \hl{are used throughout unless otherwise stated.}
% %Units note?
%%%%%%%%%%%%%%%%%%%%%%%%%%%%%%%%%%%%%%%%%%
\section{Theory}

The interaction of the electrons of an ion with the nuclear magnetic moment is described by the operator
\begin{equation}
    H_{\mathrm{hfs}} = \frac{1}{c}\vec{\mu}\cdot\vec{T}^{(1)}.
\end{equation}
Here, \hl{$c$ is the speed of light}, $\vec{\mu}$ is the operator of the nuclear magnetic moment, acting in the nuclear subspace, and electronic operator $\vec{T}^{(1)}$ is the sum of the one-electron operators $\vec{t}^{(1)}(i)$,
\begin{equation} \label{eq:T1}
   \vec{T}^{(1)} = \sum_i \vec{t}^{(1)}(i) = \sum_i\frac{\left[\vec{r}_i \times \vec{\alpha}_i \right]}{r_i^3},
\end{equation}
where $\vec{\alpha}$ are the Dirac matrices, and the point-dipole approximation is employed. We do not take into account the finite-nucleus magnetization distribution, since the corresponding correction (the Bohr-Weisskopf effect~\cite{Bohr50}) is smaller than the uncertainty associated with electron-correlations effects. It is worth noting that for heavy atoms and ions with fewer valence electrons, the theoretical accuracy is high enough that both the Bohr-Weisskopf and quantum electrodynamical \hl{(QED)} corrections to the HFS become important~\cite{Ginges17,Skripnikov24}.
The magnetic dipole HFS splitting of the state with total angular momentum $J$ can be expressed in terms of the HFS constant $A$,
\begin{equation} \label{eq:A-const}
    A =\frac{1}{c} \frac{\mu}{I} \frac{\langle J ||\vec{T}^{(1)}|| J \rangle}{\sqrt{J(J+1)(2J+1)}},
\end{equation}
where $\mu = \langle II| \mu_z|II \rangle$ is the magnetic moment of the nucleus. According to Ref.~\cite{Stone.2019}, the value $\mu = -0.2310(15)\mu_N$, where $\mu_N$ is the nuclear magneton, is used in the present calculations for $^{169}$Tm. 
The reduced matrix element of the operator $\vec{T}^{(1)}$ between the many-electron states in Eq.~\eqref{eq:A-const} is evaluated under the density matrix formalism, which allows us to express it via one-electron matrix elements~\cite{Cheung25}.

\hl{The Land\'e $g$ factor of a fine-structure level, $g_J$, defined by}
\begin{equation}
    \Delta E^{(1)} = g_J \mu_{\mathrm{B}} M_J B,
\end{equation}
\hl{where $\Delta E^{(1)}$ is the linear energy-level shift induced by the interaction with an external homogeneous magnetic field $\vec{B}$, which is assumed to be aligned with the $z$ axis, $M_J$ is the $z$ projection of the total angular momentum, and $\mu_{\mathrm{B}}$ is the Bohr magneton, %= \tfrac{1}{2c}
may serve as an estimate of the accuracy of the wave function of a state and, consequently, of its $A$ constant. 
The $g$ factors are calculated from the relation}
\begin{equation}
\Delta E^{(1)} = \langle J M_J |V_{\mathrm{m}}| J M_J \rangle,
\end{equation}
\hl{where the operator}
\begin{equation}
V_{\mathrm{m}} = \frac{1}{2} \sum_i[\vec{r}_i \times \vec{\alpha}_i] \cdot \vec{B},    
\end{equation}
\hl{represents the interaction with the magnetic field and $i$ enumerates the electrons of the ion. A more comprehensive calculation, including QED and other corrections (see, e.g., Ref.}~\cite{Spiess25}\hl{), is beyond the scope of the present paper.}

To calculate the $A$ constants and Land\'e $g$ factors in \Tm, we use the $14$-electron CI method while keeping the $54$ electrons of the Xe-like core frozen. We use a basis set of one-electron functions with orbital angular momentum $l \leqslant 4$ and principal quantum number $n \leqslant 10$, designated as $10spd\!f\!g$, and constructed as follows. First, we solve the Dirac-Hartree-Fock equations for the $[1s^2,\ldots,4f^{13},6s]$ electrons. % in the $V^N$ potential, with $N$ being the total number of electrons. 
Then, all electrons are frozen and the electron from the $6s$ shell is moved to the $5d$ shell, and the relativistic $5d$ orbitals are constructed in the frozen core
potential. The same procedure is applied to form the $6p$ orbitals, while the remaining virtual orbitals are constructed following the procedure described in Refs.~\cite{Kozlov96,Kozlov15}. The configuration space is formed by allowing single and double excitations from the $4f^{13}6s$ and $4f^{13}5d$ configurations to the whole $10spd\!f\!g$ basis set.
\hl{The calculations with the smaller $9spd\!f\!g$ and larger $11spd\!f\!g$ basis sets demonstrate convergence with respect to the basis-set size.
Higher-rank excitations can not be included due to computational limitations. However, it was recently found that they do not substantially affect the $A$ constants in neutral Tm}~\cite{Fleig23}, \hl{which presumably also holds for the ion.}
% The Hamiltonian $H$ includes magnetic part of the Breit interaction~\cite{Kozlov15}.
% We neglect the retardation and the quantum electrodynamics corrections since their contribution falls below our anticipated level of accuracy.

The interaction with the frozen core for the $A$ constants is partly accounted for by the RPA~\cite{Dzuba98}. The RPA corrections are calculated using an extended basis set, $35spd\!f\!g$, and added to the radial one-electron matrix elements of the operator $\vec{T}^{(1)}$ in the calculation of $A$ according to Eq.~\eqref{eq:A-const}. A complementary approach to account for core-valence as well as for valence-valence correlations using many-body perturbation theory was described in Ref.~\cite{Kozlov22} and employed in the Tm~\textsc{i} calculations~\cite{Bondarev24}.
%At the same time, we found corrections from the virtual orbitals accounted for by many-body perturbation theory, as in Refs.~\cite{Kozlov22,Bondarev24}, of minor importance and thus restrict the presentation to the results of pure CI calculations.
%%%%%%%%%%%%%%%%%%%%%%%%%%%%%%%%%%%%%%%%%%
\section{Results and Discussion}

In Table~\ref{tab:hfs-main}, we compare the $A$ constants calculated using the CI method, both without and with the RPA corrections (CI+RPA), with the experimental data from Refs.~\cite{Mansour89,Kebapcı24} and with the MCDHF calculations~\cite{Mansour89}.
\begin{table}[ht]
\caption{A comparison of the magnetic dipole hyperfine structure constants $A$ (in MHz) calculated using the CI and CI+RPA methods with the experimental results~\cite{Mansour89,Kebapcı24} for low-lying levels of odd parity in \Tm. The results of MCDHF calculations~\cite{Mansour89} are also shown. The level energies and configuration assignments are taken from the NIST ASD~\cite{NIST}.}
\label{tab:hfs-main}
\isPreprints{\centering}{} % Only used for preprints
%\begin{adjustwidth}{-\extralength}{0cm}
\begin{adjustwidth}{-1cm}{0cm}
\begin{tabularx}{\fulllength}{rXccccccc}
\toprule
    \multicolumn{4}{c}{} & \multicolumn{2}{c}{\textbf{Experiment}} & \multicolumn{3}{c}{\textbf{Calculation}} \\
    % \multirow{2}{*}{\multicolumn{4}{X}{}} & \multicolumn{2}{c}{Theory} & \multicolumn{2}{c}{Experiment} \\
     \cmidrule(lr){5-6} 
     \cmidrule(lr){7-9}
     $\bm{E}$\textbf{ (cm$^{-1}$)} &  \textbf{$\bm{J}$} & \textbf{Configuration} & \textbf{Term}  & \textbf{Ref.~\cite{Mansour89}} & \textbf{Ref.~\cite{Kebapcı24}}
     & \textbf{CI} & \textbf{CI+RPA} & \textbf{MCDHF~\cite{Mansour89}}  \\
\midrule
     0    & 4 & $4f^{13}(^2F_{7/2})6s_{1/2}$ & $(7/2,1/2)^o$ &               & $ -1038.3(16)$ & $-945 $ & $-1040$ &         \\
   236.95 & 3 & $4f^{13}(^2F_{7/2})6s_{1/2}$ & $(7/2,1/2)^o$ &               & $   293(6)   $ & $ 164 $ & $ 253 $ &         \\
  8769.68 & 2 & $4f^{13}(^2F_{5/2})6s_{1/2}$ & $(5/2,1/2)^o$ & $  914.9(10)^{*}$ & $   129(4)   $ & $-30  $ & $ 120 $ & $  -67 $ \\
  8957.47 & 3 & $4f^{13}(^2F_{5/2})6s_{1/2}$ & $(5/2,1/2)^o$ & $-1537.7(8) $ &                & $-1401$ & $-1506$ & $-1326 $ \\ [4pt]
 17624.65 & 2 & $4f^{13}(^2F_{7/2})5d_{3/2}$ & $(7/2,3/2)^o$ & $ -341.8(13)$ &                & $-463 $ & $-350 $ & $ -434 $ \\
 21713.74 & 3 & $4f^{13}(^2F_{7/2})5d_{3/2}$ & $(7/2,3/2)^o$ & $ -313.5(19)$ &                & $-362 $ & $-275 $ & $ -351 $ \\ 
\bottomrule
\end{tabularx}
\noindent{\footnotesize{* In Ref.~\cite{Kebapcı24}, this value was found to be incorrect and reanalyzed yielding a revised value of 125.7~MHz.}}
\end{adjustwidth}

\end{table}
As can be seen from the table, our CI results are in good agreement with the previous MCDHF calculation~\cite{Mansour89}. They are slightly closer to experiment for the levels of the $4f^{13}6s$ configuration, but somewhat further from experiment for the levels of the $4f^{13}5d$ configuration. The overall consistency between the CI and MCDHF results reflects the similarity of the two methods. Furthermore, as the CI configuration space is reduced, the corresponding predictions tend to move even closer to those of the MCDHF calculation.

For the first and four levels of the $4f^{13}6s$ configuration, the difference between the CI results and experiment is below 10\%, for the second level it is about 45\%, and for the third level the calculated $A$ constant even has the opposite sign. It is therefore instructive to examine the origin of these discrepancies more closely. The two dominant contributions to the $A$ constants of these levels arise from the diagonal density-matrix elements of the one-electron $f$ and $s$ functions. For the first and fourth levels, the total angular momentum of the $f$-hole (or 13 $f$-electrons) and $s$-electron are parallel when coupled to the total angular momentum $J$ of the level. In these cases, the $f$ and $s$ contributions to $A$ have the same sign, resulting in a large magnitude of $A$. In contrast, for the second and third levels the momenta are antiparallel, and the contributions have opposite signs, partially canceling each other and leading to a small magnitude of $A$. This cancellation makes the $A$ constants of these levels more difficult to calculate than those of the first and fourth ones. 
Taking all that into consideration, the measurement outcome for the third level, $A = 914.9(10)$~\cite{Mansour89}, is doubtful, given that the $A$ constants are known for the other three levels of the configuration. However, in the work of~\citet{Mansour89}, the $A$ constants of only two of the four \hl{$4f^{13}6s$} levels were measured and calculated, making such an analysis impossible. We note that the experimental results for the ground and first excited states obtained at COALA~\cite{Konig.2020b} agree with the data of~\citet{Kebapcı24}, but have uncertainties smaller by about two orders of magnitude~\cite{Bodnar25}.
%
% In Ref.~\cite{Kebapcı24}, this value proved to be incorrect and recalculated to 125.7~MHz.
%and equal to $-1039.125(21)$~MHz and $294.89(?)$~MHz the ground and first-excited state, respectively~\cite{Bodnar25}.
% % Hendrik Bodnar Masterarbeit ``Collinear Laser Spectroscopy of $^{155}$Tm$^+$ – $^{175}$Tm$^+$''~\cite{Bodnar25}.
%
%Meanwhile, for the first and fourth levels, where no cancellation of the dominant contributions occurs, the ratio of the calculated $A$ constants in the CI method is in excellent agreement with the experimental data. Moreover, this ratio, $A_1/A_4 \approx 0.68$, matches the ratio of the $J$-dependent factors $\tfrac{1}{\sqrt{J(J+1)(2J+1)}}$ in Eq.~\eqref{eq:A-const}. For the $A$ constants of levels belonging to more complex configurations, however, such a consideration becomes overly simplified.
%
The RPA corrections significantly reduce the discrepancy between the CI results and experiment for all levels of the $4f^{13}6s$ configuration. In particular, they yield the correct sign of $A$ for the third level, mainly by increasing the dominant one-electron $s$ contribution discussed above. 

For the levels of the $4f^{13}5d$ configuration, for which only two measurements of the $A$ constants exist, the RPA corrections are also important and reduce the discrepancy for the level at 17624.65~cm$^{-1}$. For the level at 21713.74~cm$^{-1}$, they as for the one at 17624.65~cm$^{-1}$ decrease the magnitude of $A$ obtained in the CI calculation; however, this makes it noticeably smaller than the experimental value. A possible explanation of the poorer agreement with experiment for this level, as for the one at 17624.65~cm$^{-1}$, is \hl{discussed below. 
To this end, Table}~\ref{tab:hfs-predict} \hl{presents} the calculated $A$ constants for all 11 experimentally known levels of the $4f^{13}5d$ configuration listed in the NIST ASD~\cite{NIST}. The comparison between calculated and experimental values of the Land\'e $g$ factors is also shown in the table.
\begin{table}[h]
\caption{Magnetic dipole hyperfine structure constants $A$ (in MHz) calculated using the CI and CI+RPA methods for the levels of the $4f^{13}5d$ configuration in \Tm. A comparison of the calculated and measured Land\'e $g$ factors is also given. The level energies, Land\'e $g$ factors, and configuration assignments are taken from the NIST ASD~\cite{NIST}.}
\label{tab:hfs-predict}
%\begin{adjustwidth}{-\extralength}{0cm}
\begin{adjustwidth}{-1cm}{0cm}
\centering
\begin{tabularx}{\fulllength}{rXccccccc}
\toprule
% \multicolumn{4}{c}{} & \multicolumn{1}{c}{\textbf{Experiment~\cite{NIST}}} & \multicolumn{3}{c}{\textbf{Calculation}} \\
\multicolumn{4}{c}{} & \multicolumn{2}{c}{\textbf{Land\'e }$\bm{g}$\textbf{ factor}} & \multicolumn{3}{c}{$\bm{A}$\textbf{ (MHz)}} \\
\cmidrule(lr){5-6} 
\cmidrule(lr){7-9}
$\bm{E}$\textbf{ (cm$^{-1}$)} &  $\bm{J}$ & \textbf{Configuration} & \textbf{Term}  & \textbf{Experiment~\cite{NIST}} & \textbf{CI} & \textbf{Experiment~\cite{Mansour89}} & \textbf{CI} & \textbf{CI+RPA} \\
\midrule
17624.65 & 2 & $4f^{13}(^2F_{7/2})5d_{3/2}$ & $(7/2,3/2)^o$ & 1.48  & 1.469 & $-341.8(13)$ & $-463$ & $-350$ \\ 
20228.75 & 5 & $4f^{13}(^2F_{7/2})5d_{3/2}$ & $(7/2,3/2)^o$ & 1.020 & 1.018 &              & $-312$ & $-330$ \\ 
21133.68 & 6 & $4f^{13}(^2F_{7/2})5d_{5/2}$ & $(7/2,5/2)^o$ & 1.167 & 1.166 &              & $-249$ & $-190$ \\ 
21713.74 & 3 & $4f^{13}(^2F_{7/2})5d_{3/2}$ & $(7/2,3/2)^o$ & 1.22  & 1.273 & $-313.5(19)$ & $-362$ & $-275$ \\ 
21978.77 & 2 & $4f^{13}(^2F_{7/2})5d_{5/2}$ & $(7/2,5/2)^o$ & 1.02  & 1.017 &              & $-442$ & $-574$ \\ 
22141.96 & 1 & $4f^{13}(^2F_{7/2})5d_{5/2}$ & $(7/2,5/2)^o$ & 1.325 & 1.324 &              & $-757$ & $-813$ \\ 
22457.51 & 4 & $4f^{13}(^2F_{7/2})5d_{3/2}$ & $(7/2,3/2)^o$ &       & 1.184 &              & $-318$ & $-246$ \\ 
23524.09 & 4 & $4f^{13}(^2F_{7/2})5d_{5/2}$ & $(7/2,5/2)^o$ &       & 1.033 &              & $-303$ & $-327$ \\ 
23934.73 & 3 & $4f^{13}(^2F_{7/2})5d_{5/2}$ & $(7/2,5/2)^o$ &       & 1.043 &              & $-332$ & $-395$ \\ 
24273.20 & 5 & $4f^{13}(^2F_{7/2})5d_{5/2}$ & $(7/2,5/2)^o$ & 1.18  & 1.172 &              & $-263$ & $-193$ \\ 
28874.14 & 4 & $4f^{13}(^2F_{5/2})5d_{3/2}$ & $(5/2,3/2)^o$ &       & 0.833 &              & $-492$ & $-500$ \\ 
\bottomrule
\end{tabularx}
\end{adjustwidth}
\end{table}
Our calculated $g$ factors are generally in very good agreement with the experimental data~\cite{NIST}. The only exception is the level at $21713.74$~cm$^{-1}$, where the calculated value disagrees with experiment. We attribute this discrepancy to the interaction of this level with other levels with $J=3$, in particular to the one at $23934.73$~cm$^{-1}$, which is not fully captured by our CI calculation. This, in turn, leads to a less accurate wave function and $A$ constant for that level. To examine this hypothesis, $g$ factor measurements for the other levels in Table~\ref{tab:hfs-predict} would be valuable. An experimental study of the corresponding $A$ constants would also be of interest, in order to verify the CI+RPA predictions, which show a large variation of the $A$ constants among the levels of the $4f^{13}5d$ configuration. \hl{Overall, the accuracy of the calculated $A$ constants for the levels with good agreement in the $g$ factors is expected to be similar to that for the level at 17624.65~cm$^{-1}$.}

%For these levels, the uncertainty of the $A$ constant is thus also expected to be larger than for the other levels presented. Comparison between the calculated $g$ factors and experimental data helps to estimate the accuracy of our predictions for the $A$ constants. 

%%%%%%%%%%%%%%%%%%%%%%%%%%%%%%%%%%%%%%%%%%
\section{Conclusions}

In this work, we used the CI method to calculate the magnetic dipole HFS constants $A$ and Land\'e $g$ factors for the levels of the ground $4f^{13}6s$ and first excited $4f^{13}5d$ odd configurations in \Tm. We found that the RPA corrections to the $A$ constants are essential for achieving good agreement between theory and experiment for the levels of the ground configuration, for which, owing to its relative simplicity, a more detailed analysis has been presented. The obtained results provide predictions for states where experimental data are not yet available. Our approach can also be used to calculate the electric field gradient at the site of the nucleus, which gives rise to the electric quadrupole interaction in unstable isotopes with nuclear spin $I \geqslant 1$. This information is important for planning experiments with radioactive isotopes; however, in contrast to transition rates and $A$ constants, it can not be obtained in ground-step experiments using the stable isotope $^{169}$Tm.

%%%%%%%%%%%%%%%%%%%%%%%%%%%%%%%%%%%%%%%%%%
\vspace{6pt} 
%%%%%%%%%%%%%%%%%%%%%%%%%%%%%%%%%%%%%%%%%%
%\authorcontributions{For research articles with several authors, a short paragraph specifying their individual contributions must be provided. The following statements should be used ``Conceptualization, X.X. and Y.Y.; methodology, X.X.; software, X.X.; validation, X.X., Y.Y. and Z.Z.; formal analysis, X.X.; investigation, X.X.; resources, X.X.; data curation, X.X.; writing---original draft preparation, X.X.; writing---review and editing, X.X.; visualization, X.X.; supervision, X.X.; project administration, X.X.; funding acquisition, Y.Y. All authors have read and agreed to the published version of the manuscript.'', please turn to the  \href{http://img.mdpi.org/data/contributor-role-instruction.pdf}{CRediT taxonomy} for the term explanation. Authorship must be limited to those who have contributed substantially to the work~reported.}

% \funding{Please add: ``This research received no external funding'' or ``This research was funded by NAME OF FUNDER grant number XXX.'' and  and ``The APC was funded by XXX''. Check carefully that the details given are accurate and use the standard spelling of funding agency names at \url{https://search.crossref.org/funding}, any errors may affect your future funding.}

\funding{This research received no external funding.}

% \informedconsent{Any research article describing a study involving humans should contain this statement. Please add ``Informed consent was obtained from all subjects involved in the study.'' OR ``Patient consent was waived due to REASON (please provide a detailed justification).'' OR ``Not applicable'' for studies not involving humans. You might also choose to exclude this statement if the study did not involve humans.

% Written informed consent for publication must be obtained from participating patients who can be identified (including by the patients themselves). Please state ``Written informed consent has been obtained from the patient(s) to publish this paper'' if applicable.}

\dataavailability{The original contributions presented in this study are included in the article. Further inquiries can be directed to the author.}%{We encourage all authors of articles published in MDPI journals to share their research data. In this section, please provide details regarding where data supporting reported results can be found, including links to publicly archived datasets analyzed or generated during the study. Where no new data were created, or where data is unavailable due to privacy or ethical restrictions, a statement is still required. Suggested Data Availability Statements are available in section ``MDPI Research Data Policies'' at \url{https://www.mdpi.com/ethics}.} 

\acknowledgments{\hl{I} thank Mikhail Kozlov for valuable discussions, and Hendrik Bodnar 
and Wilfried N\"ortersh\"auser 
for sharing unpublished experimental results.}

\conflictsofinterest{The author declares no conflicts of interest.} 

%%%%%%%%%%%%%%%%%%%%%%%%%%%%%%%%%%%%%%%%%%

\abbreviations{Abbreviations}{
The following abbreviations are used in this manuscript:
\\

\noindent 
\begin{tabular}{@{}ll}
HFS & Hyperfine structure \\
MCDHF & Multiconfiguration Dirac-Hartree-Fock \\
COALA & Collinear apparatus for laser spectroscopy and applied physics \\
CI & Configuration interaction \\
RPA & Random-phase approximation \\
QED & Quantum electrodynamical 
% NIST & National Institute of Standards and Technology \\
% ASD & Atomic Spectra Database \\
% TW & This work
\end{tabular}
}

%%%%%%%%%%%%%%%%%%%%%%%%%%%%%%%%%%%%%%%%%%
%% Optional
\appendixtitles{no} % Leave argument "no" if all appendix headings stay EMPTY (then no dot is printed after "Appendix A"). If the appendix sections contain a heading then change the argument to "yes".
% \appendixstart
% \appendix
% \section[\appendixname~\thesection]{}
% \subsection[\appendixname~\thesubsection]{}
% The appendix is an optional section that can contain details and data supplemental to the main text---for example, explanations of experimental details that would disrupt the flow of the main text but nonetheless remain crucial to understanding and reproducing the research shown; figures of replicates for experiments of which representative data are shown in the main text can be added here if brief, or as Supplementary Data. Mathematical proofs of results not central to the paper can be added as an appendix.

% \begin{table}[H] 
% \caption{This is a table caption.\label{tab5}}
% %\newcolumntype{C}{>{\centering\arraybackslash}X}
% \begin{tabularx}{\textwidth}{CCC}
% \toprule
% \textbf{Title 1}	& \textbf{Title 2}	& \textbf{Title 3}\\
% \midrule
% Entry 1		& Data			& Data\\
% Entry 2		& Data			& Data\\
% \bottomrule
% \end{tabularx}
% \end{table}

% \section[\appendixname~\thesection]{}
% All appendix sections must be cited in the main text. In the appendices, Figures, Tables, etc. should be labeled, starting with ``A''---e.g., Figure A1, Figure A2, etc.

%%%%%%%%%%%%%%%%%%%%%%%%%%%%%%%%%%%%%%%%%%
% \isPreprints{}{% This command is only used for ``preprints''.
%\begin{adjustwidth}{-\extralength}{0cm}

\reftitle{References}

% \nocite{*}
\bibliography{refs}

@Misc{NIST,
author = {A.~Kramida and {Yu.~Ralchenko} and
J.~Reader and {NIST ASD Team}},
HOWPUBLISHED = {{NIST Atomic Spectra Database
(ver. 5.12), [Online]. Available:
{\tt{https://physics.nist.gov/asd}} [Mon Dec 08 2025].
National Institute of Standards and Technology, Gaithersburg, MD.}},
year = {2024},
url={https://doi.org/10.18434/T4W30F}
}

@misc{Müller25,
      title={Hyperfine spectroscopy of optical-cycling transitions in singly ionized thulium}, 
      author={Patrick Müller and Andrei Tretiakov and Amanda Younes and Nicole Halawani and Paul Hamilton and Wesley C. Campbell},
      year={2025},
      eprint={2512.14885},
      archivePrefix={arXiv},
      primaryClass={physics.atom-ph},
      url={https://arxiv.org/abs/2512.14885}, 
}

@misc{Bodnar25,
  author       = {Bodnar, Hendrik and N\"ortersh\"auser, Wilfried and others},
  year         = {2025},
  note         = {Manuscript in preparation},
}

@techreport{Cheal:2834596,
      author        = "Cheal, Bradley and Rodriguez, Liss and Bai, Shiwei and
                       Blaum, Klaus and Campbell, Paul and Garcia Ruiz, Ronald and
                       Imgram, Philip and Koenig, Kristian and Lellinger, Tim and
                       Muller, Patrick and Nazarewicz, Witold and Neugart, Rainer
                       and Neyens, Gerda and Nies, Lukas and Nortershauser,
                       Wilfried and Page, Robert and Plattner, Peter and Reinhard,
                       Paul-Gerhard and Renth, Laura and Rothe, Sebastian and
                       Sanchez, Rodolfo and Schweiger, Christoph and Stegemann,
                       Simon and Stora, Thierry and Wang, Simin and Yang, Xiaofei
                       and Yordanov, Deyan",
      title         = "{Laser spectroscopy of neutron-deficient thulium
                       isotopes}",
      institution   = "CERN",
      reportNumber  = "CERN-INTC-2022-041, INTC-I-245",
      address       = "Geneva",
      year          = "2022",
      url           = "https://cds.cern.ch/record/2834596",
      note          = {\url{https://cds.cern.ch/record/2834596}}
}

@techreport{Cheal:2872390,
      author        = "Cheal, Bradley and Vazquez, Rodriguez",
      title         = "{Laser spectroscopy of neutron-deficient thulium
                       isotopes}",
      institution   = "CERN",
      reportNumber  = "CERN-INTC-2023-059, INTC-P-673",
      address       = "Geneva",
      year          = "2023",
      url           = "https://cds.cern.ch/record/2872390",
      note          = {\url{https://cds.cern.ch/record/2872390}}      
}

@techreport{Cheal:2912229,
      author        = "Cheal, Bradley and Vazquez Rodrigues, Liss and Heinke,
                       Reinhard",
      title         = "{Laser spectroscopy of neutron-deficient thulium
                       isotopes}",
      institution   = "CERN",
      reportNumber  = "CERN-INTC-2024-065, INTC-P-673-ADD-1",
      address       = "Geneva",
      year          = "2024",
      url           = "https://cds.cern.ch/record/2912229",
      note          ={\url{https://cds.cern.ch/record/2912229}}
}

@article{Cheung25,
title = {p{CI}: A parallel configuration interaction software package for high-precision atomic structure calculations},
journal = {Computer Physics Communications},
volume = {308},
pages = {109463},
year = {2025},
issn = {0010-4655},
doi = {10.1016/j.cpc.2024.109463},
url = {https://www.sciencedirect.com/science/article/pii/S0010465524003862},
author = {Charles Cheung and Mikhail G. Kozlov and Sergey G. Porsev and Marianna S. Safronova and Ilya I. Tupitsyn and Andrey I. Bondarev},
}

@article{Bondarev24,
  title = {Comparison of theory and experiment for radiative characteristics in neutral thulium},
  author = {Bondarev, Andrey I. and Tamanis, Maris and Ferber, Ruvin and Ba\ifmmode \mbox{\c{s}}\else \c{s}\fi{}ar, G\"on\"ul and Kr\"oger, Sophie and Kozlov, Mikhail G. and Fritzsche, Stephan},
  journal = {Phys. Rev. A},
  volume = {109},
  issue = {1},
  pages = {012815},
  numpages = {11},
  year = {2024},
  month = {Jan},
  publisher = {American Physical Society},
  doi = {10.1103/PhysRevA.109.012815},
  url = {https://link.aps.org/doi/10.1103/PhysRevA.109.012815}
}

@article{Spiess25,
  title = {Excited-State Magnetic Properties of Carbon-like ${\mathrm{Ca}}^{14+}$},
  author = {Spie\ss{}, Lukas J. and Chen, Shuying and Wilzewski, Alexander and Wehrheim, Malte and Gilles, Jan and Surzhykov, Andrey and Benkler, Erik and Filzinger, Melina and Steinel, Martin and Huntemann, Nils and Cheung, Charles and Porsev, Sergey G. and Bondarev, Andrey I. and Safronova, Marianna S. and Crespo L\'opez-Urrutia, Jos\'e R. and Schmidt, Piet O.},
  journal = {Phys. Rev. Lett.},
  volume = {135},
  issue = {4},
  pages = {043002},
  numpages = {7},
  year = {2025},
  month = {Jul},
  publisher = {American Physical Society},
  doi = {10.1103/p88p-brnx},
  url = {https://link.aps.org/doi/10.1103/p88p-brnx}
}

@article{Kebapcı24,
doi = {10.3847/1538-4357/ad47b7},
url = {https://doi.org/10.3847/1538-4357/ad47b7},
year = {2024},
month = {jul},
publisher = {The American Astronomical Society},
volume = {970},
number = {1},
pages = {23},
author = {Kebapcı, Taha Yusuf and Parlatan, \c{S}eyma and Sert, Sami and Öztürk, \.{I}pek Kanat and Ba\c{s}ar, Gönül and \c{S}ahin, Timur and Bilir, Sel\c{c}uk and Ferber, Ruvin and Tamanis, Maris and Kröger, Sophie},
title = {Hyperfine Structure Investigation of Singly Ionized Thulium in Fourier-transform Spectra},
journal = {The Astrophysical Journal}
}

@article{Mansour89,
title = {High-precision measurements of hyperfine structure in Tm II, N$_2^+$ and Sc II},
journal = {Nuclear Instruments and Methods in Physics Research Section B: Beam Interactions with Materials and Atoms},
volume = {40-41},
pages = {252-256},
year = {1989},
issn = {0168-583X},
doi = {10.1016/0168-583X(89)90972-5},
url = {https://www.sciencedirect.com/science/article/pii/0168583X89909725},
author = {N.B. Mansour and T.P. Dinneen and L. Young}
}

@article{Konig.2020b,
 author = {K{\"o}nig, K. and Kr{\"a}mer, J. and Geppert, C. and Imgram, P. and Maa{\ss}, B. and Ratajczyk, T. and N{\"o}rtersh{\"a}user, W.},
 year = {2020},
 title = {A new Collinear Apparatus for Laser Spectroscopy and Applied Science (COALA)},
 pages = {081301},
 volume = {91},
 number = {8},
 journal = {Rev. Sci. Instrum.},
 doi = {10.1063/5.0010903}
}

@article{Cheng85,
  title = {Ab initio calculation of $4f^N6s^2$ hyperfine structure in neutral rare-earth atoms},
  author = {Cheng, K. T. and Childs, W. J.},
  journal = {Phys. Rev. A},
  volume = {31},
  issue = {5},
  pages = {2775--2784},
  numpages = {0},
  year = {1985},
  month = {May},
  publisher = {American Physical Society},
  doi = {10.1103/PhysRevA.31.2775},
  url = {https://link.aps.org/doi/10.1103/PhysRevA.31.2775}
}

@article{DenHartog24,
doi = {10.3847/1538-4365/ad614f},
url = {https://doi.org/10.3847/1538-4365/ad614f},
year = {2024},
month = {aug},
publisher = {The American Astronomical Society},
volume = {274},
number = {1},
pages = {9},
author = {Den Hartog, E. A. and Voith, G. T. and Roederer, I. U.},
title = {Atomic Transition Probabilities for Ultraviolet and Optical Lines of Tm II},
journal = {The Astrophysical Journal Supplement Series}
}

@article{Fleig23,
  title = {Suppressed electric quadrupole moment of thulium atomic clock states},
  author = {Fleig, Timo},
  journal = {Phys. Rev. A},
  volume = {107},
  issue = {3},
  pages = {032816},
  numpages = {7},
  year = {2023},
  month = {Mar},
  publisher = {American Physical Society},
  doi = {10.1103/PhysRevA.107.032816},
  url = {https://link.aps.org/doi/10.1103/PhysRevA.107.032816}
}

@article{Kozlov15,
title = {CI-MBPT: A package of programs for relativistic atomic calculations based on a method combining configuration interaction and many-body perturbation theory},
journal = {Computer Physics Communications},
volume = {195},
pages = {199-213},
year = {2015},
issn = {0010-4655},
doi = {10.1016/j.cpc.2015.05.007},
url = {https://www.sciencedirect.com/science/article/pii/S001046551500185X},
author = {M.G. Kozlov and S.G. Porsev and M.S. Safronova and I.I. Tupitsyn}
}

@article{Kozlov22,
  title = {Combination of perturbation theory with the configuration-interaction method},
  author = {Kozlov, M. G. and Tupitsyn, I. I. and Bondarev, A. I. and Mironova, D. V.},
  journal = {Phys. Rev. A},
  volume = {105},
  issue = {5},
  pages = {052805},
  numpages = {10},
  year = {2022},
  month = {May},
  publisher = {American Physical Society},
  doi = {10.1103/PhysRevA.105.052805},
  url = {https://link.aps.org/doi/10.1103/PhysRevA.105.052805}
}

@article{Radžiūtė21,
doi = {10.3847/1538-4365/ac1ad2},
url = {https://doi.org/10.3847/1538-4365/ac1ad2},
year = {2021},
month = {nov},
publisher = {The American Astronomical Society},
volume = {257},
number = {2},
pages = {29},
author = {Radžiūtė, Laima and Gaigalas, Gediminas and Kato, Daiji and Rynkun, Pavel and Tanaka, Masaomi},
title = {Extended Calculations of Energy Levels and Transition Rates for Singly Ionized Lanthanide Elements. II. Tb-Yb},
journal = {The Astrophysical Journal Supplement Series}
}

@article{Quinet99,
title = {On the use of the Cowan’s code for atomic structure calculations in singly ionized lanthanides},
journal = {Journal of Quantitative Spectroscopy and Radiative Transfer},
volume = {62},
number = {5},
pages = {625-646},
year = {1999},
issn = {0022-4073},
doi = {10.1016/S0022-4073(98)00127-7},
url = {https://www.sciencedirect.com/science/article/pii/S0022407398001277},
author = {P. Quinet and P. Palmeri and E. Biémont},
}

@article{Parlatan22,
title = {Experimental investigation of the hyperfine structure of Tm I with Fourier transform spectroscopy, part A: In the visible wavelength range (400–700 nm)},
journal = {Journal of Quantitative Spectroscopy and Radiative Transfer},
volume = {287},
pages = {108195},
year = {2022},
issn = {0022-4073},
doi = {10.1016/j.jqsrt.2022.108195},
url = {https://www.sciencedirect.com/science/article/pii/S0022407322001303},
author = {{Parlatan}, {\c{S}}eyma and {{\"O}zt{\"u}rk}, {\.I}pek K. Kanat  and {Ba\c{s}ar}, G{\"o}n{\"u}l and {Ba\c{s}ar}, G{\"u}nay and Ferber, Ruvin  and {Kr{\"o}ger}, Sophie}
}

@article{Kebapcı22,
title = {Experimental investigation of the hyperfine structure of Tm I with Fourier transform spectroscopy part B: In the NIR wavelength range from 700 nm to 2250 nm},
journal = {Journal of Quantitative Spectroscopy and Radiative Transfer},
volume = {287},
pages = {108196},
year = {2022},
issn = {0022-4073},
doi = {10.1016/j.jqsrt.2022.108196},
url = {https://www.sciencedirect.com/science/article/pii/S0022407322001315},
author = {Taha Yusuf Kebapcı and Sami Sert and Şeyma Parlatan and İpek Kanat Öztürk and Gönül Başar and Günay Başar and Maris Tamanis and Sophie Kröger}
}

@article{Bohr50,
  title = {The Influence of Nuclear Structure on the Hyperfine Structure of Heavy Elements},
  author = {Bohr, Aage and Weisskopf, V. F.},
  journal = {Phys. Rev.},
  volume = {77},
  issue = {1},
  pages = {94--98},
  numpages = {0},
  year = {1950},
  month = {Jan},
  publisher = {American Physical Society},
  doi = {10.1103/PhysRev.77.94},
  url = {https://link.aps.org/doi/10.1103/PhysRev.77.94}
}

@article{Ginges17,
  title = {Ground-state hyperfine splitting for Rb, Cs, Fr, ${\mathrm{Ba}}^{+}$, and ${\mathrm{Ra}}^{+}$},
  author = {Ginges, J. S. M. and Volotka, A. V. and Fritzsche, S.},
  journal = {Phys. Rev. A},
  volume = {96},
  issue = {6},
  pages = {062502},
  numpages = {8},
  year = {2017},
  month = {Dec},
  publisher = {American Physical Society},
  doi = {10.1103/PhysRevA.96.062502},
  url = {https://link.aps.org/doi/10.1103/PhysRevA.96.062502}
}

@article{Skripnikov24,
  title = {Reexamination of nuclear magnetic dipole and electric quadrupole moments of polonium isotopes},
  author = {Skripnikov, Leonid V. and Barzakh, Anatoly E.},
  journal = {Phys. Rev. C},
  volume = {109},
  issue = {2},
  pages = {024315},
  numpages = {13},
  year = {2024},
  month = {Feb},
  publisher = {American Physical Society},
  doi = {10.1103/PhysRevC.109.024315},
  url = {https://link.aps.org/doi/10.1103/PhysRevC.109.024315}
}

@misc{Stone.2019,
 author = {Stone, N. J.},
 date = {2019},
 title = {Table of Recommended Nuclear Magnetic Dipole Moments: Part I - Long-lived States},
 address = {International Atomic Energy Agency (IAEA)},
 number = {INDC(NDS)--0794},
 doi = {10.61092/iaea.yjpc-cns6}
}

@article{Wyart11,
author = {Wyart, Jean-François},
title = {On the interpretation of complex atomic spectra by means of the parametric Racah–Slater method and Cowan codes},
journal = {Canadian Journal of Physics},
volume = {89},
number = {4},
pages = {451-456},
year = {2011},
doi = {10.1139/p10-112},
URL = {https://doi.org/10.1139/p10-112}
}

@BOOK{Martin78,
        author = {{Martin}, W.~C. and {Zalubas}, Romuald and {Hagan}, Lucy},
        title = "{Atomic energy levels - The rare-earth elements}",
        year = 1978,
        publisher = {Nat. Bur. Stand., U.S.},
        doi={10.6028/NBS.NSRDS.60},
        adsurl = {https://ui.adsabs.harvard.edu/abs/1978aelr.book.....M}
}

@article{Blagoev94,
title = {Lifetimes of Levels of Neutral and Singly Ionized Lanthanide Atoms},
journal = {Atomic Data and Nuclear Data Tables},
volume = {56},
number = {1},
pages = {1-40},
year = {1994},
issn = {0092-640X},
doi = {10.1006/adnd.1994.1001},
url = {https://www.sciencedirect.com/science/article/pii/S0092640X84710011},
author = {K.B. Blagoev and V.A. Komarovskii}
}

@article{Anderson96,
author = {Heidi M. Anderson and E. A. Den Hartog and J. E. Lawler},
journal = {J. Opt. Soc. Am. B},
number = {11},
pages = {2382--2391},
publisher = {Optica Publishing Group},
title = {Radiative lifetimes in Tm I and Tm II},
volume = {13},
month = {Nov},
year = {1996},
url = {https://opg.optica.org/josab/abstract.cfm?URI=josab-13-11-2382},
doi = {10.1364/JOSAB.13.002382}
}

@article{Wickliffe97,
author = {M. E. Wickliffe and J. E. Lawler},
journal = {J. Opt. Soc. Am. B},
number = {4},
pages = {737--753},
publisher = {OSA},
title = {Atomic transition probabilities for Tm I and Tm II},
volume = {14},
month = {Apr},
year = {1997},
url = {http://www.osapublishing.org/josab/abstract.cfm?URI=josab-14-4-737},
doi = {10.1364/JOSAB.14.000737}
}

@article{Rieger99,
  title = {Beam-laser lifetime measurements for some selected levels in singly ionized thulium},
  author = {Rieger, Georg and McCurdy, Michelle M. and Pinnington, Eric H.},
  journal = {Phys. Rev. A},
  volume = {60},
  issue = {5},
  pages = {4150--4152},
  numpages = {0},
  year = {1999},
  month = {Nov},
  publisher = {American Physical Society},
  doi = {10.1103/PhysRevA.60.4150},
  url = {https://link.aps.org/doi/10.1103/PhysRevA.60.4150}
}

@Article{Xu03,
author={Xu, H.-L. and Jiang, Z.-K. and Svanberg, S.},
title={Lifetime measurements in Tm I, Tm II, and Tm III by time-resolved laser spectroscopy},
journal={The European Physical Journal D - Atomic, Molecular, Optical and Plasma Physics},
year={2003},
month={Jun},
day={01},
volume={23},
number={3},
pages={323-326},
issn={1434-6079},
doi={10.1140/epjd/e2003-00055-3},
url={https://doi.org/10.1140/epjd/e2003-00055-3}
}

@article{Tian16,
    author = {Tian, Yanshan and Wang, Xinghao and Yu, Qi and Li, Yongfan and Gao, Yang and Dai, Zhenwen},
    title = "{Radiative lifetime measurements of some Tm I and Tm II levels by time-resolved laser spectroscopy}",
    journal = {Monthly Notices of the Royal Astronomical Society},
    volume = {457},
    number = {2},
    pages = {1393-1398},
    year = {2016},
    month = {02},
    issn = {0035-8711},
    doi = {10.1093/mnras/stv3015},
}

@article{Wang22,
title = {Experimental branching fractions, transition probabilities and oscillator strengths in Tm I and Tm II},
journal = {Journal of Quantitative Spectroscopy and Radiative Transfer},
volume = {280},
pages = {108091},
year = {2022},
issn = {0022-4073},
doi = {10.1016/j.jqsrt.2022.108091},
url = {https://www.sciencedirect.com/science/article/pii/S0022407322000280},
author = {Xinghao Wang and Qi Yu and Yanshan Tian and Zhiming Chen and Hongqiang Xie and Xinyi Zeng and Guangcheng Guo and Haoxiong Chang}
}

@article{Kozlov96,
doi = {10.1088/0953-4075/29/4/011},
url = {https://doi.org/10.1088/0953-4075/29/4/011},
year = {1996},
month = {feb},
publisher = {},
volume = {29},
number = {4},
pages = {689},
author = {M G Kozlov and S G Porsev and V V Flambaum},
title = {Manifestation of the nuclear anapole moment in the M1 transitions in bismuth},
journal = {Journal of Physics B: Atomic, Molecular and Optical Physics}
}

@Article{Dzuba98,
author={Dzuba, V. A.
and Flambaum, V. V.
and Kozlov, M. G.
and Porsev, S. G.},
title={Using effective operators in calculating the hyperfine structure of atoms},
journal={Journal of Experimental and Theoretical Physics},
year={1998},
month={Nov},
day={01},
volume={87},
number={5},
pages={885-890},
doi={10.1134/1.558736},
url={https://doi.org/10.1134/1.558736}
}
\end{document}